\documentclass[twoside,a4paper,11pt,reqno]{amsart}
\usepackage{amsmath}\usepackage[mathcal]{eucal}
\usepackage{times}
\numberwithin{equation}{section}

\newtheorem{theorem}{Theorem}[section]
\newtheorem{remark}[theorem]{Remark}
\numberwithin{equation}{section}

\begin{document}

\title[Relativistic Boltzmann Equation with Hard Interactions]
{Global Existence Proof for Relativistic Boltzmann Equation 
with Hard Interactions}

\author{Zhenglu Jiang}
\address{Department of Mathematics, Zhongshan University, 
Guangzhou 510275, P.~R.~China}
\email{mcsjzl@mail.sysu.edu.cn}
\thanks{This work  was supported by NSFC 10271121 and 
the Scientific Research Foundation for the Returned Overseas Chinese Scholars, 
the Ministry of Education of China, and sponsored by joint grants 
of NSFC 10511120278/10611120371 and RFBR 04-02-39026}


\date{September 29, 2007.}


\begin{abstract}
By combining the DiPerna and Lions techniques
for the nonrelativistic Boltzmann equation and the Dudy\'{n}ski and Ekiel-Je\.{z}ewska device of 
 the causality of the relativistic Boltzmann equation, 
it is shown that there exists a global mild solution to the Cauchy problem for 
the relativistic Boltzmann equation 
with the assumptions of the relativistic scattering cross section including 
some  relativistic  hard interactions and the initial data satisfying 
finite mass, energy  and entropy. This is in fact an extension of the result   
of Dudy\'{n}ski and Ekiel-Je\.{z}ewska 
to the case of the relativistic Boltzmann equation with hard interactions.
\end{abstract}

\maketitle
\vspace*{-0.7cm}
\thispagestyle{empty}

\section{Introduction}\label{intro}
We are concerned with a global existence of mild solution to the Cauchy problem for 
the relativistic Boltzmann equation with  
 the relativistic scattering cross section including  
some  relativistic  hard interactions through initial data satisfying 
finite mass, energy  and entropy. 
The relativistic Boltzmann equation 
(hereafter RBE)  is of the following dimensionless form (see \cite{gv})
\begin{equation}
\frac {\partial f}{\partial t}+\frac {{\bf p}}
{p_0}\frac{\partial f}{\partial{\bf x}}
=Q(f, f) \label{rbe}
\end{equation}
for a one-particle distribution function $ {f=f(t, {\bf x},{\bf p})}$ that depends on 
the time $t\in {\bf R}_+,$  the position ${\bf x}\in{\bf R}^3,$ 
and the momentum ${\bf p}\in{\bf R}^3,$   
  where $p_0=(1+|{\bf p}|^2)^{1/2}$ 
and $Q(f,f)$ is the relativistic collision operator 
whose structure will be addressed below.  
 Here and throughout this paper, 
 ${\bf R}_+$ represents the positive side of the real axis including its origin and 
 ${\bf R}^3$ denotes a three-dimensional Euclidean space.

The collision operator $Q$ is expressed by the difference between the gain and loss terms 
respectively defined by 
 \begin{equation}
Q^+(f,f)(t,{\bf x},{\bf p})= \int_{{\bf R}^3\times{S}^2} 
f(t,{\bf x},{\bf p}^\prime)f(t,{\bf x},{\bf p}_1^\prime)\frac{B(g, \theta)}{p_0p_{1 0}}d^3 {\bf p}_1d\Omega
\label{rbep}
\end{equation}
and 
\begin{equation}
Q^-(f,f)(t,{\bf x},{\bf p})=  \int_{{\bf R}^3\times{S}^2} 
 f(t,{\bf x},{\bf p})f(t,{\bf x},{\bf p}_1)\frac{B(g, \theta)}{p_0p_{1 0}}d^3 {\bf p}_1d\Omega.
\label{rbem}
\end{equation}
In equations (\ref{rbep}) and (\ref{rbem}),  $S^2$  
is a unit sphere surface in ${\bf R}^3,$   
$({\bf p}^\prime,{\bf p}_1^\prime)$ are dimensionless momenta  
after collision of two particles 
having precollisional dimensionless momenta $({\bf p},{\bf p}_1),$ 
$p_{10}$ is defined by $p_{10}=(1+|{\bf p}_1|^2)^{1/2}$ 
and represents the dimensionless energy of the colliding particle   
having the momentum ${\bf p}_1$  
immediately before collision of two particles,   
$B(g, \theta)$ is the collision kernel of the 
 momentum distance and scattering angle variables  $g$ and $\theta$ 
which are respectively denoted by 
\begin{equation}
g=\sqrt{|{\bf p}_1-{\bf p}|^2-|p_{10}-p_0|^2}/2 
\label{g}
\end{equation} 
and  
\begin{equation}
\theta=\arccos\{1+[(p_0-p_{10})(p_0-p_0^{\prime})
-({\bf p}-{\bf p}_1)({\bf p}-{\bf p}^{\prime})]/(2g^2)\}
\label{theta}
\end{equation}  
with  
$p_0^{\prime}=(1+|{\bf p}^{\prime}|^2)^{1/2}$ 
representing the dimensionless energy of the colliding particle   
having the momentum ${\bf p}^{\prime}$  
immediately after collision of two particles,  
and $d\Omega=\sin{\theta}d{\theta}d{\psi}$ 
is the differential of area on $S^2$ 
for any $\theta \in[0,\pi]$ and $\psi\in[0,2\pi].$ 
 
The initial data $f|_{t=0}=f_0({\bf x},{\bf p})$ in ${\bf R}^3\times{\bf R}^3$ are  
required to satisfy 
\begin{equation}
f_0\geq 0 ~\hbox{ a.e. in }{\bf R}^3\times{\bf R}^3,
\iint_{{\bf R}^3\times{\bf R}^3}f_0(1+p_0+|\ln f_0|)d^3{\bf x}d^3{\bf p}<\infty.
\label{rbec}
\end{equation}
In (\ref{rbec}),  the third term of the integral can control 
the Boltzmann entropy at an initial time while 
the two other terms of the integral, from left to right, 
 respectively represent the mass and the energy 
in the relativistic system at the initial time.  The finiteness of  all the integrals 
states that the relativistic system has  finite mass, energy and entropy 
at the initial state. 

There are many authors who have contributed to the study of 
the Cauchy problem for RBE,  e.g., Bichteler \cite{b}, Bancel \cite{bd}, 
Dudy\'{n}ski and Ekiel-Je\.{z}ewska \cite{de85} \cite{de85e} \cite{de88} \cite{de92}, Glassey and 
Strauss \cite{gs91} \cite{gs92} \cite{gs},  Andr\'easson \cite{ka}, 
Cercignani and Kremer \cite{ck}, Glassey \cite{g06}. 
Many other relevant papers and books can be found 
in the references mentioned above.  

The DiPerna and Lions techniques (see \cite{dl})  
for the nonrelativistic Boltzmann equation were originally applied 
by Dudy\'{n}ski and Ekiel-Je\.{z}ewska \cite{de92} to their proof of  
a global existence of solutions to the Cauchy problem 
for RBE with the assumptions of the relativistic scattering cross section 
excluding the relativistic hard interactions and 
 the initial data 
satisfying finite mass, energy and entropy.  
Unlike in the nonrelativistic case,    
the relativistic initial data is not required to have a finite  ``inertia'' 
since the causality of solutions 
to RBE is used by Dudy\'{n}ski and Ekiel-Je\.{z}ewska into their proof. 
Their results are correct (except the boundness of the entropy at any time 
without such an assumption as a finite ``inertia'' considered below) 
but  their assumption of the relativistic scattering cross section 
does not include the cases of the relativistic hard interactions. 
After that, a different proof was also given in \cite{j98a}  
to show a global existence of solutions to the large-data Cauchy problem 
for RBE with some relativistic hard interactions. 
In his proof, the property of the causality is not used directly in  
solving the Cauchy problem but it is assumed that 
the initial data satisfies 
\begin{equation}
f_0\geq 0 ~\hbox{ a.e. in }{\bf R}^3\times{\bf R}^3,
\iint_{{\bf R}^3\times{\bf R}^3}f_0(1+p_0|{\bf x}|^2+p_0+|\ln f_0|)d^3{\bf x}d^3{\bf p}<\infty,
\label{rbec2}
\end{equation}
i.e., finite mass, ``inertia'', energy 
and entropy.   Unlike in the nonrelativistic case,  the initial condition 
(\ref{rbec2}) indicates that 
the relativistic ``inertia'' is required 
to involve an integral of $f_0p_0|{\bf x}|^2$ 
over space and momentum variables because of the fact that 
the physically natural {\it a priori} estimates of 
the solutions to RBE are made by using 
the relativistic collision invariant $p_0({\bf x}-{\bf p}t/p_0)^2+t^2/p_0$ 
of two colliding particles immediately before and after collision 
 while those to the nonrelativistic Boltzmann equation 
result from the nonrelativistic collision invariant $({\bf x}-{\bf vt})^2.$   

The objective of this paper is to show that there exists a global
mild solutions to the large-data Cauchy problem
for RBE with some relativistic hard interactions under the condition of the initial data
 $f_0$ satisfying (\ref{rbec}), that is,
\begin{theorem}\label{th1}
Let $B(g, \theta)$ be the relativistic collision kernel of RBE (\ref{rbe}), defined above,
and $B_R$ a ball with a center at the origin and a radius $R,$
$A(g)=\int_{S^2}B(g, \theta)d\Omega .$
Assume that
\begin{equation}
B(g, \theta)\geq 0 ~\hbox{ a.e. in } [0,+\infty)\times S^2,
B(g, \theta)\in L_{loc}^1({\bf R}^3\times S^2),
\label{rbeb1}
\end{equation}
\begin{equation}
\frac{1}{p_0^2}\int_{B_R}\frac{A(g)}{p_{10}}d^3{\bf p}_1
{\rightarrow}0~ \hbox{ as  } |{\bf p}|{\rightarrow}+\infty,  ~\hbox{ for all } 
R{\in}(0,+\infty).
\label{rbeb2}
\end{equation}
Then RBE (\ref{rbe}) has a mild
or equivalently a renormalized solution $f$
through initial data $f_0$ with (\ref{rbec}),
satisfying the following properties
\begin{equation}f\in C([0,+\infty);L^1({\bf R}^3{\times}{\bf R}^3)),
\label{sc1}\end{equation}
\begin{equation}L(f)\in L^\infty([0,+\infty);L^1({\bf R}^3{\times}B_R)), 
~\hbox{ for all }   R{\in}(0,+\infty), 
\label{sc2}\end{equation}
\begin{equation}\frac{Q^+(f,f)}{1+f}\in L^1([0,T];L^1({\bf R}^3{\times}B_R)), 
~\hbox{ for all }   R, T{\in}(0,+\infty),
\label{sc3}\end{equation}
\begin{equation}
\sup\limits_{t\geq 0}\iint_{{\bf R}^3{\times}{\bf R}^3}f(1+p_0
+{\ln}f)d^3{\bf x}d^3{\bf p}<+\infty.
\label{sc4}\end{equation}
\end{theorem}
This theorem is in fact an extension of the result given by Dudy\'{n}ski and Ekiel-Je\.{z}ewska \cite{de92} to 
the relativistic system with hard interactions. 
The reason is found that both the causality of RBE 
and the conservation of mass and energy 
in the relativistic system guarantee 
the relativistic ``inertia'' 
involving an integral of  $fp_0|{\bf x}|^2$ 
over all the space and momentum variables to 
be successfully estimated at any time. 

It is clear that the condition (\ref{rbeb1}) is equivalent to the following one:  
\begin{equation}
\sigma(g, \theta)\geq 0 ~\hbox{ a.e. in } [0,+\infty)\times S^2, 
g(1+g^2)^{1/2}\sigma(g, \theta)\in L_{loc}^1([0,+\infty)\times S^2), 
\label{rbeb12}
\end{equation}
 which was first defined by Jiang \cite{j98a}.  
The assumption (\ref{rbeb2}) was originally introduced 
by Jiang  (see \cite{j97a}, \cite{j99}).  
Obviously, the relativistic assumptions  (\ref{rbeb1}) and  (\ref{rbeb2}) are similar  
to the following nonrelativistic ones adopted by DiPerna and Lions \cite{dl}: 
\begin{equation}
B({\bf z}, \omega)\geq 0 ~\hbox{ a.e. in } {\bf R}^{N}\times S^{N-1}, 
B({\bf z}, \omega)\in L_{loc}^1({\bf R}^{N}\times S^{N-1}),
\label{beb1}
\end{equation}
\begin{equation}
\frac{1}{1+|{\bf \xi}|^2}\int_{|{\bf z}-{\bf \xi}|\leq R}\tilde{A}({\bf z})d{\bf z}
{\rightarrow}0~ \hbox{ as  } |{\bf \xi}|{\rightarrow}+\infty,  ~\hbox{ for all } 
R{\in}(0,+\infty),
\label{beb2}
\end{equation}
where $B({\bf z}, \omega)$ is a function of $|{\bf z}|,$ $|({\bf z},\omega)|$ only, 
$\tilde{A}({\bf z})=\int_{S^{N-1}}B({\bf z}, \omega)d\omega .$ 
It is also easy to see that the condition (\ref{rbeb2}) includes some 
relativistic hard interactions defined as $\int_{S^{2}}B(g, \theta)d\Omega \geq Cg^2,$
where $C$ is a positive constant (see \cite{de88}).  
For example, if $B(g,\theta)=s^{\frac{1}{2}}g^{\beta+1}\sin^\gamma(\theta)$ where $\gamma>-2,$ 
$0\leq\beta<\min(2,2+\gamma),$ then $B(g,\theta)$ satisfies (\ref{rbeb1}) and (\ref{rbeb2}), 
and it is a relativistic hard interaction kernel.  
But it was assumed by Dudy\'{n}ski  and 
Ekiel-Je\.{z}ewska (see \cite{de92}) that $B(g, \theta)$ satisfies (\ref{rbeb1}) and the following condition:   
\begin{equation}
\frac{1}{p_0}\int_{B_R}\frac{A(g)}{p_{10}}d^3{\bf p}_1
{\rightarrow}0~ \hbox{ as  } |{\bf p}|{\rightarrow}+\infty,  ~\hbox{ for all } 
R{\in}(0,+\infty),
\label{rbeb2de}
\end{equation}
where $B_R$ and $A(g)$
are the same as (\ref{rbeb2});  it has been claimed in \cite{de92} that 
their assumptions of $B(g, \theta)$ exclude the relativistic hard interactions.  
In fact, since $g^2=(p_{10}p_0-{\bf p}_1{\bf p}-1)/2,$  
 it is easy to see that 
$$\int_{B_R}\frac{A(g)}{p_{10}}d^3{\bf p}_1\geq 2\pi C[p_0R^3/3-R\sqrt{1+R^2}/2+\ln(R+\sqrt{1+R^2})/2]$$ 
for $A(g)\geq C g^2,$ where $C$ is a positive constant and $R>0.$
 This implies that 
 (\ref{rbeb2de}) does not hold in the relativistic hard interaction cases.
It follows that  (\ref{rbeb2de}) is more restrictive than (\ref{rbeb2}). 

The rest of this paper is organized as follows. 
Besides the conservation laws of mass, momenta and 
energy in the relativistic system, the property that the entropy of the system 
is always a nondecreasing function of $t$ is described in section \ref{cons}.   
 Finally, in section \ref{ge}, 
the DiPerna and Lions techniques and the Dudy\'{n}ski and Ekiel-Je\.{z}ewska devices 
are successfully applied to prove the global existence of solutions to 
the Cauchy problem for RBE with hard interactions   
in $L^1$ if  the initial data satisfies finite mass, energy 
and entropy. 
The physically natural {\it a priori} estimates of the solutions 
are also shown to be bounded in any given finite time interval.

\section{Conservation Laws and Entropy}
\label{cons}
As in the nonrelativistic case, the structure of the relativistic collision operator 
maintains not only the conversation of mass, momenta 
and energy in the relativistic system,  but also 
the property that the entropy of the system does not decrease. 

Since energy and momenta of  two colliding 
particles conserve before and  after  collision, 
that is,  ${\bf p}+{\bf p}_1={\bf p}^\prime+{\bf p}_1^\prime, $ $p_0+p_{10}=p_0^{\prime}
+p_{10}^{\prime} ,$   
 it is easily proved that  
\begin{equation}
s=s^\prime,  ~~~~g=g^\prime, 
\label{tenergy}
\end{equation}
where 
$s^\prime=|p_{10}^\prime+p_0^\prime|^2-|{\bf p}_1^\prime+{\bf p}^\prime|^2,$  
$g^\prime=\sqrt{|{\bf p}_1^\prime-{\bf p}^\prime|^2-|p_{10}^\prime-p_0^\prime|^2}/2.$ 
It is also shown that 
\begin{equation}
\cos\theta=1-2[|p_0-p_0^\prime|^2-|{\bf p}-{\bf p}^\prime|^2]/(s-4). 
\label{rbea}
\end{equation}

It requires further analysis of the relativistic collision term  to 
show the conversation laws in the relativistic system.    
By using (\ref{tenergy}) and (\ref{rbea}),  
it can be easily proved that 
\begin{equation}
\int_{{\bf R}^3}\psi({\bf p})Q(\varphi, \varphi)d^3{\bf p}
=\frac{1}{4}\iint_{{\bf R}^3\times{\bf R}^3}\frac{d^3 {\bf p}_1}{p_0p_{1 0}} 
\int_{{S}^2}d\Omega B(g, \theta)
[\varphi({\bf p}^\prime)\varphi({\bf p}_1^\prime)-\varphi({\bf p})\varphi({\bf p}_1)] 
\nonumber
\end{equation}
\begin{equation}
\cdot[\psi({\bf p})+\psi({\bf p}_1)-\psi({\bf p}^\prime)-\psi(
{\bf p}_1^\prime)]
\label{rbekeq}
\end{equation}
if $ Q^\pm(\varphi, \varphi)\psi({\bf p})\in{L}^1({\bf R}^3)$
 for any given $\psi(
{\bf p})\in{L}^\infty({\bf R}^3) $ and 
every given $\varphi({\bf p})\in{L}^1({\bf R}^3). $  
It follows from (\ref{rbekeq}) that $\int_{{\bf R}^3}
{\bar \psi}Q(f, f)d^3{\bf p}=0$ if 
$f=f(t,{\bf x}, {\bf p})$ is a distributional solution to RBE (\ref{rbe}) 
such that $\int_{{\bf R}^3}{\bar \psi}Q(f, f)d^3{\bf p}<+\infty$ for almost all $t$ and ${\bf x}$ and 
${\bar \psi}={\bar b_0}+{\bf b}
{\bf p}+c_0p_0, $ where
 ${\bar b_0}\in{\bf R}, 
{\bf b}\in{\bf R}^3, c_0\in{\bf R}.$
Furthermore,  it is at least formally found that 
$\iint_{{\bf R}^3\times{\bf R}^3}{\bar \psi}fd^3{\bf x}d^3{\bf p}$ 
 is independent of $t$  for any distributional solution $f$ to RBE (\ref{rbe}).
This yields the conservation of mass, momentum and kinetic energy of the 
relativistic system.  

It is well known that the nonrelativistic Boltzmann equation has the conservation of 
the integral of $f({\bf x}-{\bf v}t)^2$ over all the space and velocity variables 
besides the conservation of mass, momentum and kinetic energy of the 
nonrelativistic system. This is because $({\bf x}-{\bf v}t)^2$ is an invariant of two 
nonrelativistic colliding particles immediately before and after  collision. In the relativistic case,  
although $p_0({\bf x}-t{\bf p}/p_0)^2+t^2/p_0$ is a relativistic invariant of two 
colliding particles immediately before and after collision, 
the integral of $f[p_0({\bf x}-t{\bf p}/p_0)^2+t^2/p_0]$ 
over all the space and momentum variables changes with $t.$ In fact, 
by multiplying RBE (\ref{rbe}) by $p_0({\bf x}-t{\bf p}/p_0)^2+t^2/p_0$  
and integrating by parts over ${\bf x}$ and ${\bf p},$ it is easy to see that 
\begin{equation}
\frac{d}{dt}\iint_{{\bf R}^3{\times}{\bf R}^3}
f[p_0({\bf x}-t{\bf p}/p_0)^2+t^2/p_0]d^3{\bf x}d^3{\bf p}
\nonumber 
\end{equation}
\begin{equation}
=\iint_{{\bf R}^3{\times}{\bf R}^3}
f\left(\frac {\partial }{\partial t}+\frac {{\bf p}}
{p_0}\frac{\partial }{\partial{\bf x}}
\right)[p_0({\bf x}-t{\bf p}/p_0)^2+t^2/p_0]d^3{\bf x}d^3{\bf p}
\label{dinvariant}
\end{equation}
and hence 
\begin{equation}
\frac{d}{dt}\iint_{{\bf R}^3{\times}{\bf R}^3}
f[p_0({\bf x}-t{\bf p}/p_0)^2+t^2/p_0]d^3{\bf x}d^3{\bf p}
=2t\iint_{{\bf R}^3{\times}{\bf R}^3}
f/p_0d^3{\bf x}d^3{\bf p}.
\label{dinvariant01}
\end{equation}  
which yields the estimate of 
the integral $\iint_{{\bf R}^3{\times}{\bf R}^3}
f[p_0({\bf x}-t{\bf p}/p_0)^2+t^2/p_0]d^3{\bf x}d^3{\bf p}$ 
under the assumption of (\ref{rbec2}). 
This is why the assumption (\ref{rbec2}) was really made by Jiang \cite{j98a} before.  
Fortunately,  
it can be easily known from (\ref{dinvariant01}) that 
\begin{equation}
\sup\limits_{0\leq t\leq T}\iint_{{\bf R}^3{\times}{\bf R}^3}
fp_0({\bf x}-t{\bf p}/p_0)^2d^3{\bf x}d^3{\bf p}
\leq \iint_{{\bf R}^3{\times}{\bf R}^3}
f_0(p_0|{\bf x}|^2+T^2)d^3{\bf x}d^3{\bf p},
\label{dinvariantentropy}
\end{equation}  
which is very useful to the estimate of the relativistic 
entropy integral considered below. 

By (\ref{dinvariantentropy}), the desired estimate of  
$\iint_{{\bf R}^3{\times}{\bf R}^3}
fp_0|{\bf x}|^2d^3{\bf x}d^3{\bf p}$ under the assumption of  (\ref{rbec2}) 
can be also made successfully. To show this estimate, it requires   
the following identity  
\begin{equation}
\frac{d}{dt}\iint_{{\bf R}^3{\times}{\bf R}^3}
fp_0|{\bf x}|^2d^3{\bf x}d^3{\bf p}
=2\iint_{{\bf R}^3{\times}{\bf R}^3}
f{\bf x}{\bf p}d^3{\bf x}d^3{\bf p}
\label{dinvariant02}
\end{equation}
derived by multiplying RBE (\ref{rbe}) by $p_0|{\bf x}|^2$ 
and integrating by parts over ${\bf x}$ and ${\bf p},$ and hence 
\begin{equation}
\frac{d}{dt}\iint_{{\bf R}^3{\times}{\bf R}^3}
fp_0|{\bf x}|^2d^3{\bf x}d^3{\bf p}
\leq \iint_{{\bf R}^3{\times}{\bf R}^3}
fp_0|{\bf x}|^2d^3{\bf x}d^3{\bf p}+\iint_{{\bf R}^3{\times}{\bf R}^3}
fp_0d^3{\bf x}d^3{\bf p},
\label{dinvariant03}
\end{equation} 
which yields the following inequality 
\begin{equation}
\sup\limits_{0\leq t\leq T}\iint_{{\bf R}^3{\times}{\bf R}^3}
fp_0|{\bf x}|^2d^3{\bf x}d^3{\bf p}
\leq e^T \iint_{{\bf R}^3{\times}{\bf R}^3}
f_0p_0(1+|{\bf x}|^2)d^3{\bf x}d^3{\bf p} 
\label{dinvariant04}
\end{equation} 
for any given $T>0$ by multiplying the two sides of (\ref{dinvariant03}) by $e^{-t}$ 
and  using the conservation of the mass of the  relativistic system.  
The inequality given by (\ref{dinvariant04}) illustrates   
that the relativistic ``inertia'' of $fp_0|{\bf x}|^2$ over all the space and 
momentum variables is at any time controlled  by both mass and ``inertia'' 
at the initial state of the relativistic system. 
The above analysis also dedicates that 
 the conservation of mass and energy 
 guarantees  
the relativistic ``inertia'' 
involving an integral of  $fp_0|{\bf x}|^2$ 
over all the space and momentum variables to 
be successfully estimated in the relativistic system at any time. 

The physically natural estimates of solutions 
to RBE (\ref{rbe}) require not only 
the relativistic conservation laws but also the property that 
the entropy is always a nondecreasing function of $t$ 
in the relativistic system.  To show this property of the relativistic entropy, 
the relativistic entropy identity has to be first considered 
as in the nonrelativistic case.
It is easy to at least formally deduce the following entropy identity 
\begin{equation}
\frac{d}{dt}\iint_{{\bf R}^3\times{\bf R}^3}f\ln fd^3{\bf x}d^3{\bf p}
+\frac{1}{4p_0}\iint_{{\bf R}^3\times{\bf R}^3}\frac{d^3 {\bf p}_1}{p_{1 0}} 
\int_{{S}^2}d\Omega B(g, \theta)
\nonumber
\end{equation}
\begin{equation}
\cdot[f(t,{\bf x},{\bf p}^\prime)f(t,{\bf x},{\bf p}_1^\prime)
-f(t,{\bf x},{\bf p})f(t,{\bf x},{\bf p}_1)]
\ln\left[\frac{f(t,{\bf x},{\bf p}^\prime)f(t,{\bf x},{\bf p}_1^\prime)}
{f(t,{\bf x},{\bf p})f(t,{\bf x},{\bf p}_1)}\right]=0 
\label{rbeei}
\end{equation}
by multiplying RBE (\ref{rbe}) by $1+\ln f,$ 
integrating over ${\bf x}$ and ${\bf p}$ 
and using (\ref{rbekeq}).  In general,  for convenience, put 
$H(t)=\iint_{{\bf R}^3\times{\bf R}^3}f\ln fd^3{\bf x}d^3{\bf p},$ 
and $H(t)$ is called H-function.  
Boltzmann's entropy is usually defined by $-H(t).$ 
The second term in (\ref{rbeei}) is nonnegative and so 
 $H(t)$ is a nonincreasing function of $t.$ 
This means that 
the entropy of the relativistic system does not decrease. 
This property allows the desired estimate of the relativistic entropy 
to be derived from the Cauchy problem for RBE. 

In fact, the entropy can be controlled by the integral 
$\iint_{{\bf R}^3\times{\bf R}^3}f|\ln f|d^3{\bf x}d^3{\bf p}$ 
for any nonnegative solution to RBE (\ref{rbe}) 
 and so  it is natural to make the considered estimate 
of the integral instead of the entropy.  
  Notice that 
\begin{equation}
\iint_{{\bf R}^3\times{\bf R}^3}f|\ln f|d^3{\bf x}d^3{\bf p}=\iint_{{\bf R}^3\times{\bf R}^3}f\ln fd^3{\bf x}d^3{\bf p}
+2\iint_{f\leq 1}f|\ln f|d^3{\bf x}d^3{\bf p}
\nonumber
\end{equation}
\begin{equation}
\hspace*{0.1cm}\leq \iint_{{\bf R}^3\times{\bf R}^3}f\ln fd^3{\bf x}d^3{\bf p}
+2\iint_{{\bf R}^3\times{\bf R}^3}f[p_0({\bf x}-t{\bf p}/p_0)^2+p_0]d^3{\bf x}d^3{\bf p} 
\nonumber
\end{equation}
\begin{equation}
\hspace*{2cm}+2\iint_{f\leq \exp(-|{\bf x}-t{\bf p}/p_0|^2-p_0)}f\ln(1/f)d^3{\bf x}d^3{\bf p} 
\nonumber
\end{equation}
\begin{equation}
\hspace*{0.1cm}\leq \iint_{{\bf R}^3\times{\bf R}^3}f\ln fd^3{\bf x}d^3{\bf p}
+2\iint_{{\bf R}^3\times{\bf R}^3}f[p_0({\bf x}-t{\bf p}/p_0)^2+p_0]d^3{\bf x}d^3{\bf p}
+C_1
\label{enineq}
\end{equation}
where $C_1$ is some positive constant independent of $f.$
By  using (\ref{dinvariantentropy}), (\ref{rbeei}) and (\ref{enineq}), it can be deduced that 
\begin{eqnarray}
\sup\limits_{0\leq t\leq T}\left[\iint_{{\bf R}^3\times{\bf R}^3}f|\ln f|d^3{\bf x}d^3{\bf p}\right]
\hspace*{4cm}\nonumber\\ 
\leq \iint_{{\bf R}^3\times{\bf R}^3}f_0[2T^2+2p_0(1+|{\bf x}|^2)+|\ln f_0|]d^3{\bf x}d^3{\bf p}+C_1.
\label{rbeest}
\end{eqnarray}
This implies that the boundness of the entropy at any time might not be guaranteed 
without such an assumption as the finite initial ``inertia" mentioned above.  

It is worth mentioning that much other properties of RBE (\ref{rbe})  
can be found  from the book of Cercignani and Kremer \cite{ck}.

\section{Proof of Global Existence}
\label{ge} 
In order to prove Theorem \ref{th1}, 
both the collision kernel and the initial data 
have to be first  truncated and regularized by using  the same approximation scheme 
as given by DiPerna and Lions \cite{dl} in the nonrelativistic case. 
The collision kernel $B(g, \theta)$ of RBE (\ref{rbe}) 
can be truncated  to obtain  
$B_n(g, \theta)\in L^\infty\cap L^1({\bf R}^3;L^1(S^2))$ such that
\begin{equation}
\iint_{B_R\times S^2}d^3 {\bf p}d\Omega|B_n(g, \theta)-B(g, \theta)|\to 0  
\label{arbekc}
\end{equation}
uniformly in $\{{\bf p}_1: |{\bf p}_1|\leq k\}$ as $n\to +\infty$ for all $ R,k\in (0,+\infty).$ 
Then it leads to the problem of solving the approximate equation    
\begin{equation}
\frac {\partial f^n}{\partial t}+\frac {{\bf p}}
{p_0}\frac {\partial f^n}{\partial{\bf x}}
=\tilde{Q}_n(f^n, f^n) ~\hbox{ in } 
(0,\infty)\times{\bf R}^3\times{\bf R}^3. 
\label{arbe}
\end{equation}
Here and below, $\tilde{Q}_n$ is defined by  
$\tilde{Q}_n(\varphi, \varphi)
=(1+\frac{1}{n}\int_{{\bf R}^3}|\varphi | d^3{\bf p})^{-1}Q_n(\varphi,\varphi)$ and 
\begin{equation}
Q_n(\varphi,\varphi)=\frac{1}{p_0}\int_{{\bf R}^3}\frac{d^3 {\bf p}_1}{p_{1 0}} 
\int_{{S}^2}d\Omega 
[\varphi({\bf p}^\prime)\varphi({\bf p}_1^\prime)-\varphi({\bf p})\varphi({\bf p}_1)]
B_n(g, \theta).  
\label{arbek}
\end{equation}

It follows from (\ref{arbek}) that for all $\varphi,\psi \in L^\infty([0, +\infty)\times{\bf R}^3{\times}{\bf R}^3) 
 \cap L^1({\bf R}^3{\times}{\bf R}^3),$
\begin{equation}
||\tilde{Q}_n(\varphi, \varphi)||_{L^\infty([0, +\infty)\times{\bf R}^3{\times}{\bf R}^3)}
\leq C_n||\varphi ||_{L^\infty([0, +\infty)\times{\bf R}^3{\times}{\bf R}^3)},
\label{arbek1}
\end{equation}
\begin{equation}
||\tilde{Q}_n(\varphi, \varphi)||_{L^1({\bf R}^3{\times}{\bf R}^3)}
\leq C_n||\varphi ||_{L^1({\bf R}^3{\times}{\bf R}^3)},
\label{arbek2}
\end{equation}
\begin{equation}
||\tilde{Q}_n(\varphi, \varphi)-\tilde{Q}_n(\psi, \psi)||_{L^1({\bf R}^3{\times}{\bf R}^3)}
\leq C_n||\varphi- \psi ||_{L^1({\bf R}^3{\times}{\bf R}^3)}, 
\label{arbek3}
\end{equation}
here and below everywhere, $C_n$ is a nonnegative constant independent of $\varphi$ and $\psi.$ 

By following DiPerna and Lions \cite{dl}, 
the initial data $f_0$ can be first truncated and regularized to get   
a sequence of nonnegative functions $f_0^n\in D({\bf R}^3\times{\bf R}^3)$  such that  
\begin{equation}
\iint_{{\bf R}^3\times{\bf R}^3}d^3{\bf x}d^3{\bf p}
|f_0-f_0^n|(1+p_0|{\bf x}|^2+p_0)\to 0 \hbox{ as } n\to +\infty,
\end{equation}
\begin{equation}
\iint_{{\bf R}^3\times{\bf R}^3}d^3{\bf x}d^3{\bf p}
f_0^n|\ln f_0^n|\leq C \hbox{ independent of } n.
\end{equation} 

Then there exists a unique nonnegative distributional solution $f_m^n=f_m^n(t, {\bf x}, {\bf p})$ 
to the problem of the approximate equation (\ref{arbe}) 
with the initial data $f_{m,0}^n\equiv f_0^n1_{B_m}({\bf x})$  
 for any given ball $B_m\equiv \{{\bf x}: |{\bf x}|<m\}.$  
It can be also easily proved that  $\tilde{Q}_n(f_m^n, f_m^n)
\in L_{loc}^1({\bf R}^3\times{\bf R}^3)$ and that  $f_m^n$ satisfies the following properties: 
\begin{equation}
 0\leq f_m^n{\in}L^{\infty}{\cap}L^1((0, T){\times}{\bf R}^3{\times}{\bf R}^3)~~({\forall}T<
+{\infty}),
\label{prop1} 
\end{equation}  
\begin{equation}
 f_m^n
(t, {\bf x}, 
{\bf p}){\in}
C([0, +{\infty});L^1({\bf R}^3{\times}{\bf R}^3)). 
\label{prop2}
\end{equation}

Let $\tilde{L}_n$ be denoted by 
\begin{equation}
\tilde{L}_n(\varphi)=(1+\frac{1}{n}\int_{{\bf R}^3}|\varphi | d^3{\bf p})^{-1}\frac{1}{p_0}\int_{{\bf R}^3}\frac{d^3 {\bf p}_1}{p_{1 0}}
\int_{{S}^2}d\Omega \varphi({\bf p}_1)B_n(g, \theta).
\label{arbekl}
\end{equation}
Put $\tilde{Q}_n^-(\varphi,\varphi)=\varphi({\bf p})\tilde{L}_n(\varphi)$ and 
$\tilde{Q}_n^+(\varphi,\varphi)=\tilde{Q}_n(\varphi,\varphi)-\tilde{Q}_n^-(\varphi,\varphi).$   
It is then obvious to see that 
\begin{equation}
\tilde{Q}^+_n(f_m^n, f_m^n), \tilde{Q}^-_n(f_m^n, f_m^n)
{\in}L^1_{loc}((0, +{\infty}){\times}{\bf R}^3{\times}{\bf R}^3). 
\label{prop3}
\end{equation}
By using (\ref{dinvariant04}) and (\ref{rbeest})  
and with the help of Gronwall's inequality, it can be further found that 
\begin{equation}
 \sup\limits_{t\geq 0}{{\int}{\int}}_{{\bf R}^3{\times}{\bf R}^3}
 f_m^n(1+p_0+{\ln}f_m^n
)d^3{\bf x}d^3{\bf p}{\leq}C_0,
\label{prop41}
\end{equation}\begin{equation}
 \sup\limits_{0\leq t\leq T}{{\int}{\int}}_{{\bf R}^3{\times}{\bf R}^3}
 f_m^n(p_0|{\bf x}|^2+|{\ln}f_m^n|)d^3{\bf x}d^3{\bf p}{\leq}C_{Tm}.
\label{prop4}
\end{equation}
It also follows by (\ref{rbeei}) that 
\begin{equation}
 \frac{1}{4}{\int}^{+\infty}_0{\int}_{{\bf R}^3} \left\{(1+{\int}_{{\bf R}^3} f_m^n
d^3 {\bf p})^{-1}\iiint_{{\bf R}^3\times{\bf R}^3\times{S}^2}\frac{d^3 {\bf p}_1d^3 {\bf p}d\Omega }
{p_{1 0}p_0} \right.B_n(g, \theta)
\nonumber
\end{equation}
\begin{equation}
\cdot\left.(f_m^{n{\prime}}f^{n{\prime}}_{m1}-f_m^nf^n_{m1})
{\ln}\left(\frac{f_m^{n{\prime}}f^{n{\prime}}_{m1}}
{f_m^nf^n_{m1}}\right)\right\}d{\sigma}d^3 {\bf x}
 {\leq}C_{Tm}.
\label{prop5}
\end{equation}
In (\ref{prop41}), $C_0$ is a positive constant  which is only dependent of  
 $f_0.$
In (\ref{prop4}) and (\ref{prop5}), $C_{Tm}$ is a positive constant  dependent of $m,$ 
 $f_0$ and $T$ except of $n.$  It can be easily deduced from (\ref{prop41}) and (\ref{prop4}) that 
$\{f_m^n\}^{\infty}_{n=1}$ is 
  weakly compact in $L^1((0, T){\times}
{\bf R}^3{\times}{\bf R}^3)$ for $T\in (0,+{\infty})$ when $m$ is fixed. 
Thus it may be assumed 
without loss of generality that $ f_m^n$ converges weakly in 
$L^1((0, T){\times}{\bf R}^3{\times}{\bf R}^3)$ to 
$f_m{\in}L^1_{loc}([0, +{\infty}){\times}{\bf R}^3{\times}{\bf R}^3)$
as $n{\rightarrow}+{\infty}$
 for all $T<+{\infty}$ when $m$ is fixed. Obviously, $f_m\geq 0$ and  
 $f_m|_{t=0}=f_{m,0}$ 
for almost every $({\bf x}, {\bf p}){\in}{\bf R}^3{\times}
{\bf R}^3$ where $f_{m,0}\equiv f_01_{B_m}({\bf x}).$  

It can be also proven that for any fixed $m,T, R \in (0,+{\infty}),$  
$\{\tilde{Q}_n^\pm(f_m^n,f_m^n)/(1+f_m^n)\}^{\infty}_{n=1} $ 
are weakly compact subsets of  
$L^1((0, T){\times}{\bf R}^3{\times}B_R).$  
It further  follows that  $f_m$ is a global mind 
solution to RBE (\ref{rbe}) with the initial data $f_{m,0},$ satisfying 
\begin{equation}f_m\in C([0,+\infty);L^1({\bf R}^3{\times}{\bf R}^3)),
\label{asc1}\end{equation}
\begin{equation}L(f_m)\in L^\infty([0,+\infty);L^1({\bf R}^3{\times}B_R)), 
~\hbox{ for all }  R{\in}(0,+\infty), 
\label{asc2}\end{equation}
\begin{equation}\frac{Q^+(f_m,f_m)}{1+f_m}\in L^1([0,T];L^1({\bf R}^3{\times}B_R)), 
~\hbox{ for all }  R, T{\in}(0,+\infty),
\label{asc3}\end{equation}
\begin{equation}
\sup\limits_{m\geq 1,t\geq 0}\iint_{{\bf R}^3{\times}{\bf R}^3}f_m(1+p_0+{\ln}f_m
)d^3{\bf x}d^3{\bf p}<+\infty,
\label{asc4}\end{equation}
 by analyzing step by step 
 the relaxation of the normalization and construction of subsolutions and supersolutions 
with a similar device to that 
 given by DiPerna and Lions \cite{dl}.  This analysis not only allows for  the relations among 
three different types of solutions to RBE (\ref{rbe}) (see \cite{j98a}) but also requires 
the momentum-averaged compactness of the transport operator
of RBE (\ref{rbe}) (see \cite{gl} or \cite{j97}). Here, (\ref{asc4}) is derived from (\ref{prop41}). 

Below  is a modification of the devices of Dudy\'{n}ski and Ekiel-Je\.{z}ewska \cite{de92}. 
By using both the causality and the uniqueness of solution 
to the approximate relativistic Boltzmann equation (\ref{arbe}), 
it is easy to see that if $n$ is fixed, $f_m^n$ is 
convergent as $m\to \infty$ for almost every  $(t, {\bf x}, {\bf p}).$  
Put $f^n=\lim\limits_{m\to\infty}f_m^n.$ Then $f^n$ is a unique global distributional solution 
to the approximate equation (\ref{arbe}) through $f_0^n.$ It can be also found that 
$\{f^n\}_{n=1}^\infty$  is weakly compact in  
 $L^1((0,T)\times B_m\times{\bf R^3})$ for any given $T>0$ and $m>0.$ 
It may be assumed without loss of generality that $f^n$  
converges weakly in $L^1((0,T)\times{\bf R^3}\times{\bf R^3})$ to $f$  for any given $T>0.$
It follows that $f_m$ converges to $f$ as $m\to \infty$ for almost every  $(t, {\bf x}, {\bf p}).$
Hence $f$ is a global 
mild solution to RBE (\ref{rbe}) through $f_0$.  
By (\ref{asc1}), (\ref{asc2}), (\ref{asc3}) and (\ref{asc4}),   
it can be also shown that $f$ satisfies 
(\ref{sc1}), (\ref{sc2}), (\ref{sc3}) and (\ref{sc4}). 
This completes the proof of Theorem \ref{th1}. 

\begin{remark}
The content of this paper advances that contained in references  \cite{j97} \cite{j97a} \cite{j98a} \cite{j99}.
One advantage is to employ the core new estimates (\ref{dinvariant04}) and (\ref{rbeest}) to obtain a unique 
nonnegative distributional solution $f_m^n$ 
to the problem of the approximate equation (\ref{arbe}) 
with a class of initial data which is more natural than the ones considered
previously by Dudy\'{n}ski and Ekiel-Je\.{z}ewska.  
Another is to use the assumptions (\ref{rbeb1}) and (\ref{rbeb2}) of  the relativistic collision kernel 
with some relativistic hard interactions to show that the Cauchy problem of 
 RBE (\ref{rbe}) has a global mild solution 
on the condition of  the finite initial physically natural bounds 
excluding the finite initial ``inertia''. 
\end{remark}

\noindent {\bf Acknowledgement.}~~
The author would like to thank the referees of this paper for their valuable comments and suggestions.

\end{document}